\documentclass[10pt, a4paper]{article}

\usepackage[final]{lrec-coling2024} 
\usepackage{multirow}
\usepackage{rotating}
\usepackage{color, colortbl}
\usepackage{multicol}
\usepackage{tablefootnote}  
\usepackage{booktabs}
\usepackage{amsmath,wasysym}
\definecolor{Gray}{gray}{0.9}
\makeatletter
\def\@mb@citenamelist{cite,citep,citet,citealp,citealt,citepalias,citetalias}
\makeatother
\newcites{languageresource}{~}
\usepackage{graphicx}
\usepackage{tabularx}
\usepackage{soul}
\usepackage{titlesec}
\usepackage{xstring}
\usepackage{color}
\usepackage{subfig}
\usepackage{caption}
\usepackage{MnSymbol}
\usepackage{amsfonts}
\usepackage{xcolor}
\usepackage{hyperref}
\definecolor{darkblue}{rgb}{0, 0, 0.5}
\hypersetup{colorlinks=true, citecolor=darkblue, linkcolor=darkblue, urlcolor=darkblue}

\name{Yejin Kim$^{1,2}$, Scott Rome$^{2}$, Kevin Foley$^{2}$, Mayur Nankani$^{2}$,\\
{\bf \large Rimon Melamed$^{1}$, Javier Morales$^{2}$, Abhay Yadav$^{2}$, Maria Peifer$^{2}$,}\\
{\bf \large Sardar Hamidian$^{2\dagger}$ and H. Howie Huang$^{1\dagger}$}
\thanks{$^\dagger$ These authors are corresponding authors.}}

\address{$^{1}$GraphLab, George Washington University \\
$^{2}$Applied AI Research, Comcast\\
         \small{\textit{\{yeijnjenny,rmelamed,howie\}@gwu.edu}}\\
         \small{\textit{\{scott\_rome,kevin\_foley,javier\_moralesdelgado,abhay\_yadav3,maria\_peifer,sardar\_hamidian\}@comcast.com}}}

\title{\textbf{Improving Content Recommendation: Knowledge Graph-Based Semantic Contrastive Learning for Diversity and Cold-Start Users}}

\abstract{
Addressing the challenges related to data sparsity, cold-start problems, and diversity in recommendation systems is both crucial and demanding.
Many current solutions leverage knowledge graphs to tackle these issues by combining both item-based and user-item collaborative signals.
A common trend in these approaches focuses on improving ranking performance at the cost of escalating model complexity, reducing diversity, and complicating the task.
It is essential to provide recommendations that are both personalized and diverse, rather than solely relying on achieving high rank-based performance, such as Click-through rate, Recall, etc.
In this paper, we propose a hybrid multi-task learning approach, training on user-item and item-item interactions.
We apply item-based contrastive learning on descriptive text, sampling positive and negative pairs based on item metadata. 
Our approach allows the model to better understand the relationships between entities within the knowledge graph by utilizing semantic information from text.
It leads to more accurate, relevant, and diverse user recommendations and a benefit that extends even to cold-start users who have few interactions with items.
We perform extensive experiments on two widely used datasets to validate the effectiveness of our approach.
Our findings demonstrate that jointly training user-item interactions and item-based signals using synopsis text is highly effective. 
Furthermore, our results provide evidence that item-based contrastive learning enhances the quality of entity embeddings, as indicated by metrics such as uniformity and alignment.
\\ \newline \Keywords{knowledge graph neural network, semantic contrastive learning, multi-task learning, cold-start problems, diversity recommendation, content recommendation}
}

\begin{document}
\maketitleabstract

\section{Introduction}
The prevalence of content platform services has made a vast collection of media content accessible to consumers. 
As a result, personalized recommendation systems have improved the user experience by understanding their preference and identifying relevant content.
Furthermore, accurate and diverse recommendations have emerged as essential elements within recommendation systems.

\begin{figure}[t]
\centering
	\includegraphics[width=1.0\linewidth]{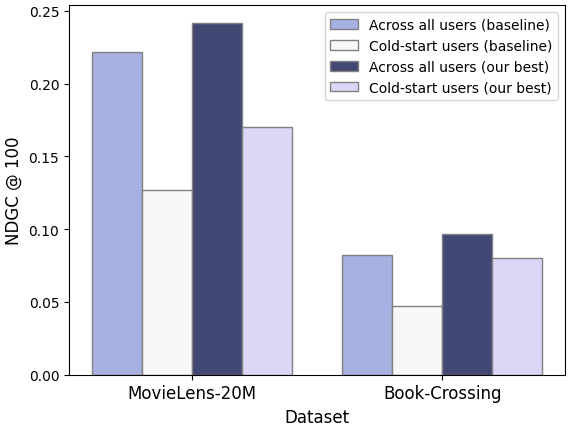}
\caption{Performance Decline in Cold-Start Scenario. In order to establish a baseline for comparing our model, we take into account the KGCN \cite{wang2019knowledge1}, a pivotal component of KG recommendation models. `Cold-start users' refers to the subgroup of test users positioned in the bottom 1\%, signifying those with the most limited user-content interactions.
}
\label{fig:intro}
\end{figure}

Collaborative Filtering (CF) \cite{miyahara2000collaborative, linden2003amazon, hofmann2004latent, su2006collaborative, jeunen2022embarrassingly} has been a popular approach in recommendation systems where the system predicts the preferences of a user based on the preferences of similar users. CF-based methods capture user-item interactions by specific operations such as inner products \cite{sarwar2001item, wang2017joint} or neural networks \cite{he2017neural}. In the context of this study, we refer to the term `item' as `content' moving forward. However, there are several problems associated with a CF-based model. It requires a significant amount of user-content data in order to prevent data sparsity problems \cite{srebro2004maximum, rennie2005fast, takacs2008investigation}. 
Moreover, it has a cold-start problem \cite{yu2004probabilistic, adomavicius2005toward} which means that it struggles to make recommendations for new users with no existing data. To tackle these issues, researchers have turned to exploring the addition of user and content features \cite{cheng2016wide} and linking them together \cite{yu2014personalized,zhang2016collaborative,zhao2017meta,huang2018improving,wang2018ripplenet,wang2018dkn}. 

Recent research has focused on constructing a Knowledge Graph (KG) that links contents to their attributes and incorporating these graphs into collaborative interactions. \citet{wang2019knowledge1} introduced the Knowledge Graph Convolutional Network (KGCN) as an extension to non-spectral GCN techniques, specifically for recommender systems. Several variant models \cite{wang2019knowledge2,wang2019kgat,wang2020ckan,wang2021learning} utilize new machine learning techniques, such as adding latent variables or network layers. 
These studies have been proposed with a specific focus on enhancing the performance of rank-based recommendations.
While performance improvement is undoubtedly important, overemphasizing ranking-based measures as an evaluation metric can lead to overlooking qualitative factors such as diversity and critical situations such as cold-start.
Figure \ref{fig:intro} visually demonstrates the decline in performance in the cold-start scenario. Notably, in the domains of both movie and book recommendations, a more significant reduction in the Normalized Discounted Cumulative Gain (NDCG)@100 is evident in the baseline model when compared to our model.

In this paper, we introduce a novel recommendation approach that leverages semantic text information using KGCN to enhance performance from multiple perspectives. Our method uses text embeddings obtained from Pre-trained Language Models (PLMs) to understand the relationships between entities in KG. 
PLMs \cite{kenton2019bert, liu2019roberta} have addressed numerous limitations of simple one-hot vector encoding. Additionally, the utilization of contrastive learning with PLMs has exhibited superior performance in sentence representation \cite{radford2019language, raffel2019exploring, oh2022don}. However, this approach has not been explored in enhancing the training of KG-based recommendations. Our study investigates the effectiveness of semantic embeddings which are learned through content-based contrastive learning. We apply the proposed method to the MovieLens-20M\cite{harper2015movielens} and Book-Crossing\footnote{\url{https://grouplens.org/datasets/book-crossing/}} datasets. The evaluation of our proposed approach is conducted along three dimensions: (1) Recommendation performance using Click-Through Rate (CTR) such as Area Under the Curve (AUC) and F1-score, along with ranking metrics including Recall and NDCG in both standard and cold-start scenarios, (2)  Assessing the level of personalization and diversity of recommended contents by Inter- and Intra- list diversity metrics, and (3) Measuring the quality of embeddings through metrics that assess uniformity and alignment. The experimental evaluation shows that our proposed method outperforms baseline approaches by taking into account users' personalized interests and diversity in the recommendation process, with an even more substantial performance boost for cold-start users. In addition, the results from the experiments confirm that content-based contrastive learning effectively regularizes content embeddings to achieve better uniformity and alignment of positive pairs in both movie and book domains. Our contributions in this paper are as follows:

\begin{itemize}
    \item 
{We propose a multi-task learning approach that utilizes user-content and content-content interactions to enrich content recommendations from various perspectives.}
    \item 
{To the best of our knowledge, we are the first to introduce content-based contrastive learning with semantic text for better KG entity relationships, yielding accurate, diverse user recommendations.}
    \item 
{Our empirical assessment affirms the effectiveness of our approach in improving recommendation performance, personalization, diversity, and the quality of content embeddings, extending its utility to demanding scenarios such as those involving cold-start users.}
    \item 
{We provide insights into the importance of comprehensive evaluation and analysis in obtaining valuable recommendations, emphasizing the need to consider multiple factors beyond rank-based performance metrics.}
\end{itemize}

\begin{figure*}
\centering 
\subfloat[Multi-task model]{\includegraphics[width=.39\linewidth]
{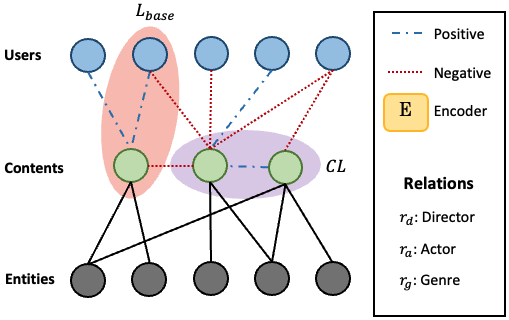}}\hspace{5pt}
\subfloat[Content-based contrastive loss ($CL$)]{\includegraphics[width=.59\linewidth]{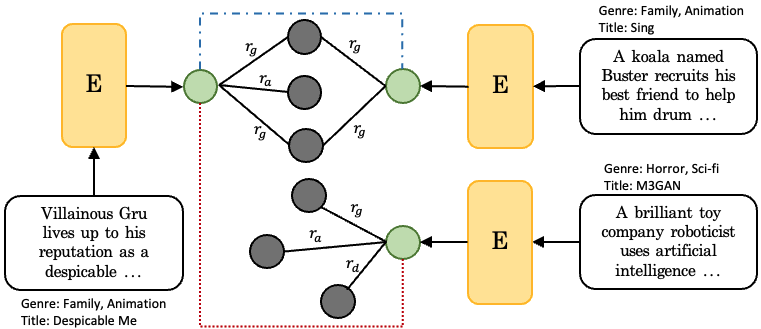}}
\caption{Overview of the Proposed Model. (a) Illustration of a knowledge graph structure and a model with multiple objectives. The loss from user-content interactions is labeled $L_{base}$, and the content-based contrastive loss is $CL$. (b) Detailed process of our proposed objective function $CL$. To generate initial content node embeddings, a pre-trained language model encodes the synopsis of each content. The positive and negative pairs are selected for each content based on their genre or title metadata (as outlined in Section \ref{encoder}).} 
\label{fig:model}
\end{figure*}

\begin{figure*}
    \centering
    \includegraphics[width=1.0\textwidth]{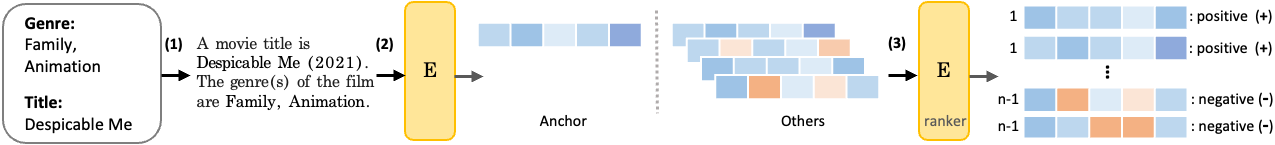}
    \caption{Sampling Positive/Negative Pairs Using Cross-encoder for $CL$}
    \label{fig:sampling}
\end{figure*}

\section{Related Works}
\subsection{Pre-trained Language Model}
Pre-trained Language Models (PLMs) have shown outstanding performance on many Natural Language Processing (NLP) tasks related to understanding the semantic meaning of sentences and context. Fine-tuned Transformer-based \cite{vaswani2017attention} language models such as Bidirectional Encoder Representations from Transformers (BERT) \cite{kenton2019bert} or Universal sentence encoder \cite{cer2018universal} models have been successfully applied in various domains including text classification \cite{sun2019fine}, named entity recognition \cite{kim2022weakly}, question answering \cite{yang2019end}, and semantic search \cite{deshmukh2020ir,esteva2020co, rome2022learning}. These models are capable of capturing more intricate relationships between words and sentences, including contextual information. 

\paragraph{PLMs in Recommendation} Fine-tuned PLMs have gained significant attention in recommendation tasks. \citet{penha2020does} worked on the usage of BERT in conversational recommender systems. They highlighted the effectiveness of pre-trained transformer models, particularly BERT, in language modeling and their capacity to implicitly retain factual information for recommendation purposes. \citet{spillo2023combining} employed a fine-tuned BERT to encode content-based representations and subsequently integrated them into graph convolutional networks to capture collaborative and structured item data. The pre-trained BERT achieved outstanding breakthroughs in the ranking task as well which is critical in the recommendation task. The BECR model, introduced by \citet{yang2021composite}, presented a re-ranking scheme that combines deep contextual token interactions with traditional lexical term-matching features. \citet{wu2021empowering} proposed an end-to-end multi-task learning framework for product ranking with fine-tuned domain-specific BERT to address the issue of vocabulary mismatch between queries and products. \citet{RAUF2024103686}
introduced BCE4ZSR which utilizes fine-tuned transformer models to distill knowledge from a larger encoder to enhance news recommendations.


\subsection{Contrastive Learning}
 The objective of Contrastive Learning (CL) is to obtain high-quality embeddings by bringing pairs of samples with similar meanings together in close proximity while pushing apart dissimilar ones \cite{hadsell2006dimensionality}.  A crucial topic in contrastive learning revolves around the construction of pairs of similar (positive) data points $(x_i, x_i^{+}
 )$. Obtaining positive pairs is more difficult than dissimilar (negative) pairs because negative examples can be drawn randomly from each training batch. In NLP, researchers devise schemes to augment data sets to produce more positive samples, such as using back-translation, paraphrasing, or generating new sentences using generative models. Similar objectives of contrastive learning have been examined in various contexts \cite{henderson2017efficient, gillick2019learning,karpukhin2020dense,gao2021simcse}. In Equation (\ref{eq1}), $z_i$ represents the embeddings of sentence $x_i$, $z_i^{+}$ denotes positive samples that are similar to $z_i$. Additionally, $t$ refers to a temperature hyperparameter suggested by \citet{DBLP:conf/cvpr/WuXYL18}. The loss of N in-batch negative samples \cite{chen2017sampling} can be expressed using the following equation. 
 {
 \begin{equation}\label{eq1}      
-log\frac{e^{({sim(z_{i},z_{i}^{^{+}})/t})}}{\sum_{j=1}^{N}e^{({sim(z_{i},z_{j}^{^{+}})/t})}} 
\end{equation}
}
\paragraph{CL in Recommendation} \citet{park2022exploiting} proposed contrastive learning for a music recommendation system based on Siamese neural networks \cite{koch2015siamese} as follows: 
 {\small
 \begin{equation}\label{eq2}      
 L=yD^2 + (1-y)max(margin-D, 0)^2
\end{equation}
}
where $y$ represents the label assigned to an item pair and $D$ stands for the distance between the items. When presented with a pair of items labeled as $y = 1$, the distance is reduced during training as $L = D^2$. Conversely, for a pair of items labeled as $y = 0$, the distance gets close to the $margin$ value according to $max(margin-D, 0)^2$. \citet{ma2023enhancing} applied $K$-means clustering algorithms to produce prototypes for user and item clusters, aiming to minimize the disparity between these prototypes using a contrastive objective detailed in Equation (\ref{eq1}). The work by \citet{li2023hkgcl} introduces graph contrastive learnings across multiple and separate domains to explore both domain-shared and domain-specific preference features for target users to improve recommendation performance, especially in sparse interaction scenarios.

\section{Proposed Methods}
Our main goal is to explore how content-based contrastive learning using semantic text affects recommendation performance when it is trained jointly with conventional collaborative loss. In Figure \ref{fig:model} (a), our proposed model is depicted, where we denote the conventional collaborative loss as $L_{base}$ and our proposed loss as $CL$. Further details about the model will be discussed in this section.
\subsection{Sampling Strategy}\label{encoder}

Figure \ref{fig:sampling} illustrates the positive/negative sampling process to train via $CL$. (1) We transform the metadata of the genre or the genre and title into sentences through a simple template. When we use the genre information solely, we create a sentence by utilizing the template: \textit{``The genre(s) of the (film|book) is/are [genre\_list]."} Likewise, when using the genre and title information together, we follow the template: \textit{``A (movie|book) title is (movie\_title|book\_title) (released\_year). The genre(s) of the (film|book) is/are [genre\_list]."} (2) The transformed sentences are input to $E$ and assigned embedding values. (3) The embeddings are passed to the cross-encoder, $E_{ranker}$, which takes in a query and the embeddings for processing. The query is one of the embeddings (\textit{Anchor} in Figure \ref{fig:sampling}) and the cross-encoder ranks all the other embeddings according to how similar they are to the query. This process is performed for each embedding. The positive and negative sets are determined by selecting the top-$n$ and bottom $n$ samples, respectively.

\subsection{Content-based Contrastive Loss}
To optimize the recommendation model, we adopt conventional user-content interaction loss \cite{wang2019knowledge1}. The assumption is that interactions between the user and the content should be given high scores for positive preferences and low scores for negative preferences. The loss function is as follows:
{\small
\begin{equation}\label{eq3}
  \begin{aligned}
    L_{base} = \sum_{u \in U} \bigg\{\sum_{c:y_{uc} = 1}\mathcal{F}(y_{uc}, \hat{y_{uc}}) \\ - \sum_{i=1}^{S^{u}} \mathbb{E}_{c_i \sim p(c_i)} \mathcal{F}(y_{uc}, \hat{y_{uc}})\bigg\}
  \end{aligned}
\end{equation}
}
where $\hat{y_{uc}}$ is predicted probability, $\mathcal{F}$ is cross-entropy loss, and $p$ is a user's negative preference sampling distribution. $S^{u}$ is the number of negative samples for user $u$, which is equal to $|\{c:y_{uc}=1\}|$. 

Figure \ref{fig:model} (b) illustrates a detailed process of our proposed content-based contrastive loss ($CL$). The loss function is represented by Equation (\ref{eq4}). $C$ is a collection of sentences transformed from metadata of the genre or the genre and title. The positive and negative sets from Section \ref{encoder} are denoted by $P \subset C$ and $N \subset C$, respectively. $P$ contains samples that are similar to the content while $N$ contains dissimilar samples. $h_c$ denotes the representation of the transformed sentences $c \in C$. Let $h_c^{+}$ and $h_c^{-}$ indicate positive and negative sample of $h_c$. Inner product, $sim(h_c, h_c^{+})= h_c \cdot h_c^{+}$, is used for our similarity function.

{\small
\begin{equation}\label{eq4}
  \begin{aligned}
CL = \sum_{c \in C} \bigg\{ \sum_{c^+ \in P}\mathcal{F}(y_{cc^{+}}, sim(h_{c}, h_c^{+}))\\-\sum_{c^- \in N}\mathcal{F}(y_{cc^{-}}, sim(h_{c}, h_c^{-}))  \bigg\} 
  \end{aligned}
\end{equation}
}
{
\begin{equation}\label{eq5}      
L=\gamma L_{base}+(1-\gamma)CL+\lambda||\Theta||_{2}^{2}
\end{equation}
}

Finally, we obtain the loss function to train Equation (\ref{eq3}) and (\ref{eq4}) jointly in Equation (\ref{eq5}), where $\gamma$ is a balancing hyper-parameter of losses. $\Theta=\{U,V,R,W_i,b_i,\forall i \in \{0, 1, \alpha\}\}$ is the model parameter set, $U$, $V$, and $R$ are the embedding tables for all users, nodes (contents and attribute entities), and relations, respectively. $W_i$ and $b_i$ denote weights of $i^{th}$-layer and $\alpha$ layer represents an additional layer specifically dedicated to content embeddings. The last term is an $L_2$-regularizer parameterized by $\lambda$ on $\Theta$.

\subsection{Training} \label{training}
Our approach follows the training procedure of the Knowledge Graph Convolutional Network (KGCN) \cite{wang2019knowledge1}. KGCN follows the identical node feature update mechanism as spatial GCN. The difference lies in the aggregation of importance scores for each relation across all users. In the equation $H^{(l+1)} = \sigma(H^{(l)} W^{(l)}+b^{(l)})$, the variables $H^{(l)}$, $W^{(l)}$, and $b^{(l)}$ represent the node features, trainable weights, and bias at layer $l$, respectively. In a standard GCN, $H^{(l)}$ is derived through the aggregation, either by summation or concatenation \cite{hamilton2017inductive} of a node's inherent features with those of its neighboring nodes. In contrast, KGCN introduces an additional step where an importance score is applied to each neighbor node's features before aggregation. KCGN is best optimized through the use of the summation aggregator while we adopt the concatenation aggregator since initializing node features for contents and attribute entities differs in our method. Our content embeddings are derived from the synopsis representation using the encoder $E$, while other entity embeddings are initialized randomly. Concatenation ensures that both features are considered when updating the weights \cite{poursafaei2022strong}. We represent the initial embedding of contents as $h_c=MLP(E(x_c))$, where $x_c$ corresponds to the synopsis text of each content, $E$ refers to an encoder and $MLP$ is a multilayer perception. As the output representation of $E$ varies in dimensions depending on the encoders used, we utilize an additional $MLP$ layer to align the embedding dimensions with those of other node embeddings. Note that the $MLP$ layer only processes content embedding and not other attribute entity embeddings. 
\begin{figure}[t]
\centering
	\includegraphics[width=1.0\linewidth]{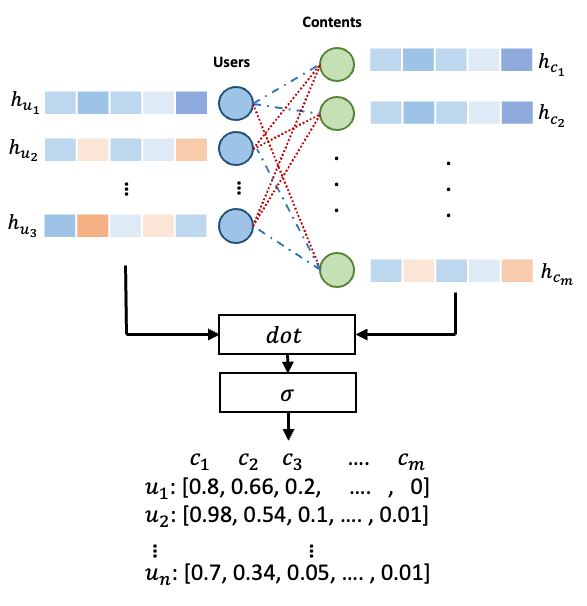}
\caption{Inference Using the Trained Contents and Users Embedding. The process $dot$ signifies the inner product operation, while $\sigma$ denotes the sigmoid transformation. Each user embedding, indexed by $n$, undergoes an inner product operation with each content embedding, indexed by $m$, followed by applying the sigmoid transformation. This procedure enables the model to rank all content for each user.
}
\label{fig:infer}
\end{figure}
\subsection{Inference} 
Figure \ref{fig:infer} illustrates the sequence of steps in the inference process. After training the graph using KGCN as detailed in Section \ref{training}, embeddings are generated for users, content, attribute entities, and relations. Notably, attribute entity embeddings and relation embeddings are not utilized in this work. We compute the inner product between each user and all content embeddings, followed by applying a sigmoid transformation. Consequently, we can rank all content items based on individual users, facilitating personalized content recommendations.

\section{Experiments}
In this section, we demonstrate the efficacy of the proposed method under two different scenarios, movie and book recommendations.  
\subsection{Experimental Setup}\label{data}
\begin{table}[hbt!]
\caption{Basic statistics and hyper-parameter settings for the two datasets (u-c: \# of user-content interactions, c-c: \# of content-content interactions by leveraging genre metadata or both genre and title metadata respectively, $K$: neighbor sampling size, $\eta$: learning rate, $d$: dimension of embeddings, $\lambda$: $L_2$-regularizer, $\gamma$: balancing parameters for losses
weight).}
    \centering
     \resizebox{0.99\columnwidth}{!}{
    \begin{tabular}{c|c|c}
    \hline\toprule
    
        
        & \textbf{MovieLens-20M}& \textbf{Book-Crossing}\\ \hline\toprule
        
        \# users& 138,159& 17,860\\
        \# contents&16,954&14,967 \\
        u-c &13,501,622&139,746 \\
        c-c&336,176 / 305,172& 221,617 / 223,140\\
        \# entities &102,569& 77,903\\
        \# relations &32&25 \\
        \# KG triples &499,474& 151,500\\\hline
        $K$ &4&4\\
        \# layers &2&1\\
        optimizer &Adam&Adam\\
        $\eta$ &2e-2&2e-4\\
        $d$  &32&64\\
        $\lambda$ & 1e-7&2e-5\\
        $\gamma$ &8e-1&8e-1\\
        batch size&65,536 (ours: 30,000) & 256\\\hline
        Text source & TMDB\footref{tmdb} & Goodreads\footref{good}, Google Books\footref{google}\\

    \hline\toprule
    \end{tabular}
    }
\label{table:data}
\end{table}

\begin{table*}[hbt!]
\caption{Performance of the AUC and F1-score were evaluated and compared against the baseline model. The highest performance is denoted in bold and the second best results are underlined. Statistical significance was assessed using t-tests; significance levels are denoted by **$p < 0.001$, *$p < 0.05$.}
    \centering
     \resizebox{0.97\textwidth}{!}
    {
    \begin{tabular}{c|c|ccc|ccc}
    \hline\toprule
    
    \multicolumn{1}{l|}{ } & \multicolumn{1}{l|}{ } & \multicolumn{3}{c|}{\textbf{Without Semantic Text}}& \multicolumn{3}{c}{\textbf{With Semantic Text}}        \\  
        \textbf{Dataset} & \textbf{Metric} &  $L_{base}$ (Baseline) & \textbf{$+ CL_{genre}$} & \textbf{$+ CL_{title+genre}$} & \textbf{\hspace{0.3cm}$L_{base}$}\hspace{0.5cm} & \textbf{$+ CL_{genre}$} & \textbf{$+ CL_{title+genre}$}  \\ \hline\toprule
        \multirow{4}{*}{Movie} & \textbf{AUC} &  0.9761 & \hspace{0.05cm}0.9773** & 0.9775** & 0.9763 & \hspace{0.3cm}\underline{0.9777}** & \textbf{0.9780}**\hspace{0.1cm} \\
          & $p$-value & - & \hspace{0.05cm}0.0007\hspace{0.3cm} & 1.96e-04 & 0.7290 & \hspace{0.25cm}0.0004\hspace{0.2cm} & 5.33e-05 \\\cline{2-8}
         & \textbf{F1} & 0.9290 & \hspace{0.05cm}0.9311*\hspace{0.2cm} & 0.9319** & 0.9284 & \hspace{0.3cm}\underline{0.9320}** & \textbf{0.9324}**\hspace{0.1cm} \\
        & $p$-value & - & \hspace{0.05cm}0.0029\hspace{0.3cm} & 1.21e-04 & 0.7330 & \hspace{0.25cm}0.0002\hspace{0.2cm} & 6.96e-05 \\ \hline\toprule
        
        \multirow{4}{*}{Book} & \textbf{AUC} &  0.6923 & 0.6911\hspace{0.25cm} & 0.6946\hspace{0.2cm} & \hspace{0.2cm}\textbf{0.7019}** & \hspace{0.35cm}\underline{0.7010}** & 0.7009**\hspace{0.05cm} \\
          & $p$-value & - & 0.6217\hspace{0.25cm} & 0.1411\hspace{0.2cm} & \hspace{0.25cm}2.39e-05 & \hspace{0.1cm}0.0008 & 0.0004\hspace{0.3cm} \\\cline{2-8}
         & \textbf{F1} & \textbf{0.6370} & 0.6336\hspace{0.25cm} & \underline{0.6362}\hspace{0.2cm} & \underline{0.6362} & \hspace{0.1cm}0.6360 & 0.6354\hspace{0.25cm} \\
        & $p$-value & - & 0.7930\hspace{0.25cm} & 0.1100\hspace{0.2cm} & 0.6011 & \hspace{0.1cm}0.6992 & 0.5824\hspace{0.25cm} \\
    \hline\toprule
    \end{tabular}
    }
\label{table:auc}
\end{table*}

    
\paragraph{Dataset} We use the MovieLens-20M~\cite{harper2015movielens} and Book-Crossing\footnote{\url{https://grouplens.org/datasets/book-crossing/}} datasets. The MovieLens-20M dataset consists of approximately 20 million explicit ratings (ranging from 1 to 5) contributed by users on the MovieLens website. Book-Crossing dataset contains 1 million ratings (ranging from 1 to 10). We follow \citet{wang2019knowledge1} when constructing a Knowledge Graph (KG), utilizing Satori by Microsoft. The basic statistics of the two datasets and KGs are presented in Table \ref{table:data}. 

We obtain 16,954 movie synopses from the TMDB database\footnote{\label{tmdb}\url{https://www.themoviedb.org}}. This process entails using the IDs mapped from MovieLens to the corresponding IDs in TMDB which is given by the MovieLens dataset. We attempt to gather synopses for 16,954 contents in our knowledge graph, but 169 of them are missing because they are not available in TMDB. Given the absence of metadata in the Book-Crossing dataset, we leverage ISBN data associated with each book entry to retrieve metadata from Goodreads\footnote{\label{good}\url{https://www.goodreads.com}}\cite{wan2018item, wan2019fine} and Google Books\footnote{\label{google}\url{https://books.google.com}}. Out of a total of 14,967 books, this methodology resulted in the successful acquisition of 11,156 genre information and 10,294 synopses.
\paragraph{Encoders} We utilize two pre-trained language models (Encoder $E$ in Figure \ref{fig:model} and Figure \ref{fig:sampling}) to represent text and one cross-encoder model (Encoder $E_{ranker}$ in Figure \ref{fig:sampling}) \cite{reimers2019sentence} for ranking the contents. We use $BERT_{base}$ \cite{kenton2019bert} for $E$ in Figure \ref{fig:model} to embed the initial input of each node's synopsis. $E$ in Figure \ref{fig:sampling} uses \textit{multi-qa-mpnet-base-dot-v1}\footnote{\url{https://www.sbert.net/docs/pretrained\_models.html}} to encode verbalized metadata. $E_{ranker}$ uses \textit{ms-marco-MiniLM-L-12-v2}\footnote{\url{https://www.sbert.net/docs/pretrained\_cross-encoders.html}} to rank the metadata embeddings according to similarity. We opt for the \textit{multi-qa-mpnet-base-dot-v1} model due to its proven superiority in embedding for semantic search. This model is able to provide optimized metadata embeddings when inputting to the cross-encoder. The cross-encoder model, \textit{ms-marco-MiniLM-L-12-v2} is fine-tuned on MS MARCO Passage Retrieval dataset~\cite{bajaj2016ms}. This particular model is highly robust in ranking text based on their similarity, as it has been fine-tuned using a dataset that comprises real user queries and their relevant text from the Bing search engine.

\paragraph{Hyper-parameters} The hyper-parameters besides batch size remain the same as those used in the original learning process. Unlike the baseline model, which utilizes a batch size of 65,536, we take a batch size of 30,000 for the movie domain. This adjustment is necessary to optimize training for both loss functions due to the significantly lower number of content-content interactions compared to user-content.

\begin{table*}[hbt!]
\caption{Performance of Recall@K and NDCG@K of the proposed model compared to the baseline model. The highest performance is denoted in bold and the second best results are underlined.}
    \centering
     \resizebox{0.99\textwidth}{!}
    {
    \begin{tabular}{c|l|ccccc|ccccc}
    \hline\toprule
    & \multicolumn{1}{l|}{ } & \multicolumn{5}{c|}{\textbf{Recall@K}} & \multicolumn{5}{c}{\textbf{NDCG@K}} \\ 
    
         & \textbf{Model} & \textbf{5} & \textbf{10} & \textbf{20} & \textbf{50} & \textbf{100}  & \textbf{5} & \textbf{10} & \textbf{20} & \textbf{50} & \textbf{100}\\ \hline\toprule
        & \multicolumn{1}{l|}{ } & \multicolumn{10}{c}{\textbf{Without Semantic Text}}\\ \hline\toprule
        \multirow{3}{*}{\begin{turn}{270}Movie\end{turn}} & $L_{base}$ (Baseline)& 0.0734 & 0.1228 & 0.1967 & 0.3214 & 0.4581 
        &0.0677	&0.0970	&0.1312	&0.1811	&0.2216\\
        & $+ CL_{genre}$ &0.0659&	0.1199	&0.1999	&0.3337&	0.4679 
        &0.0681&	0.0972	&0.1327&	0.1824	&0.2244\\
        & $+ CL_{title+genre}$ &0.0683	&0.1237&	0.1976&	\underline{0.3400}	&\textbf{0.4738}
        &0.0706	&0.1006	&0.1343	&0.1866	&0.2286\\ \hline\toprule
        
        \multirow{3}{*}{\begin{turn}{270}Book\end{turn}} & $L_{base}$ (Baseline)& \cellcolor{Gray}0.0619 & \cellcolor{Gray}0.0741 & \cellcolor{Gray}0.0926  & \cellcolor{Gray}0.1217 & \cellcolor{Gray}0.1585 
        & \cellcolor{Gray}0.0586 & \cellcolor{Gray}0.0641 & \cellcolor{Gray}0.0690  & \cellcolor{Gray}0.076 & \cellcolor{Gray}0.0825\\
        & $+ CL_{genre}$ &\cellcolor{Gray}0.0636&	\cellcolor{Gray}0.0778	&\cellcolor{Gray}0.0973	&\cellcolor{Gray}0.1317&	\cellcolor{Gray}0.1677 
        &\cellcolor{Gray}0.0630&	\cellcolor{Gray}0.0688	&\cellcolor{Gray}0.0732&	\cellcolor{Gray}0.0811	&\cellcolor{Gray}0.0873\\
        & $+ CL_{title+genre}$ &\cellcolor{Gray}\textbf{0.0774} &\cellcolor{Gray}\textbf{0.0927}&	\cellcolor{Gray}\textbf{0.1115}&	\cellcolor{Gray}\underline{0.1433}	&\cellcolor{Gray}0.1784
        &\cellcolor{Gray}0.0596	&\cellcolor{Gray}0.0643	&\cellcolor{Gray}0.0703	&\cellcolor{Gray}0.0793	&\cellcolor{Gray}0.0846\\ \hline\toprule
        
        & \multicolumn{1}{l|}{ } & \multicolumn{10}{c}{\textbf{With Semantic Text}}\\ \hline\toprule
        
        \multirow{3}{*}{\begin{turn}{270}Movie\end{turn}} & $L_{base}$ (Baseline) & \underline{0.0754} & 0.1222 & 0.2016 & 0.3260 & 0.4599 
        &0.0753	&0.1036	&0.1363	&0.1834	&0.2265\\
        & $+ CL_{genre}$ 	&\textbf{0.0755}	&\underline{0.1312}	&\underline{0.2082}	&0.3357	&\underline{0.4727}
        &\underline{0.0810}&	\underline{0.1097}	&\underline{0.1405}	&\underline{0.1911}	&\underline{0.2324} \\
        & $+ CL_{title+genre}$ &\underline{0.0754}&	\textbf{0.1325}&	\textbf{0.2136}&	\textbf{0.3418}&	0.4714
        &	\textbf{0.0882}	&\textbf{0.1154}	&\textbf{0.1498}	&\textbf{0.1983}	&\textbf{0.2419}\\ \hline\toprule

        \multirow{3}{*}{\begin{turn}{270}Book\end{turn}} & $L_{base}$ (Baseline)  & \cellcolor{Gray}0.0425 & \cellcolor{Gray}0.0645 & \cellcolor{Gray}0.0808 & \cellcolor{Gray}0.1245 & \cellcolor{Gray}0.1685
        &\cellcolor{Gray}0.0679	&\cellcolor{Gray}0.0742	&\cellcolor{Gray}0.0794	&\cellcolor{Gray}0.0871	&\cellcolor{Gray}0.0938\\
        & $+ CL_{genre}$ 	&\cellcolor{Gray}\underline{0.0730}	&\cellcolor{Gray}\underline{0.0865}	&\cellcolor{Gray}0.1032	&\cellcolor{Gray}\textbf{0.1468}	&\cellcolor{Gray}\underline{0.1823}
        &\cellcolor{Gray}\underline{0.0684}&	\cellcolor{Gray}\underline{0.0756}	&\cellcolor{Gray}\underline{0.0809}	&\cellcolor{Gray}\textbf{0.0904}	&\cellcolor{Gray}\textbf{0.0968} \\
        & $+ CL_{title+genre}$ &\cellcolor{Gray}0.0675&	\cellcolor{Gray}0.0856&	\cellcolor{Gray}\underline{0.1041}&	\cellcolor{Gray}0.1431&	\cellcolor{Gray}\textbf{0.1933}
        &	\cellcolor{Gray}\textbf{0.0744}	&\cellcolor{Gray}\textbf{0.0789}	&\cellcolor{Gray}\textbf{0.0820}	&\cellcolor{Gray}\underline{0.0892}	&\cellcolor{Gray}\underline{0.0961}\\
        
    \hline\toprule
    \end{tabular}
    }
\label{table:recall_ndcg}
\end{table*}

\subsection{Evaluation Metrics}

Our evaluation is conducted through Fidelity Jurity\footnote{\url{https://github.com/fidelity/jurity}}, an authoritative research library that provides recommender system evaluations and encompasses three different experimental scenarios. First, we evaluate Click-Through Rate (CTR) prediction by measuring interactions in the test set using the Area Under the Curve (AUC) and F1-score metrics. Second, we use the trained model to select top-$K$ contents with the highest predicted probability for each user in the test set in order to make top-$K$ recommendations. The evaluation is performed through Recall@$K$ and Normalized Discounted Cumulative Gain (NDCG)@$K$. Furthermore, we assess NDCG performance under a cold-start scenario to analyze how it is impacted by varying levels of user activity. Third, we assess the diversity among users' recommended content lists using the Inter-list diversity@$K$ metric: 
{
\begin{equation}\label{eq6}
\frac{\sum_{i,j,\{u_i, u_j\} \in I}(cosine\_distance(R_{u_i}, R_{u_j}))}{|I|}
\end{equation}
}
where $U$ denote the set of $N$ unique users, $u_i, u_j \in U$ denote the $i^{th}$ and $j^{th}$ user in the user set, $i, j \in \{1, ..., N\}$. $R_{u_i}$ is the binary indicator vector representing provided recommendations for $u_i$. $I$ is the set of all unique user pairs, $\forall i < j,\{u_i, u_j\}\in I$.
Additionally, we calculate the average pairwise cosine distances of the contents recommended to a user based on the content embeddings of our trained model. Then the results from all users are averaged as the metric Intra-list diversity@$K$: 
{
 \begin{equation}\label{eq7}      
\frac{1}{N}\sum_{i=1}^{N}\frac{\sum_{p,q,\{p, q\} \in I^{u_i}}(cosine\_distance(v_p^{u_i}, v_q^{u_i}))}{|I^{u_i}|} 
\end{equation}
}
where $v_p^{u_i}, v_q^{u_i}$ are the content features of the $p^{th}$ and $q^{th}$ content in the list of contents recommended to $u_i$, $p, q \in \{0, 1, ..., k-1\}$. $I^u$ is the set of all unique pairs of content indices for $u_i, \forall p < q, \{p, q\} \in I^{u_i}$.
Lastly, we verify the quality of our embeddings by using alignment and uniformity metrics \cite{wang2020understanding}. The alignment metric serves to gauge the proximity of features within positive pairs, while the uniformity metric evaluates the distribution of normalized features on the hypersphere.

\begin{figure}[t]
\centering
	\includegraphics[width=1.0\linewidth]{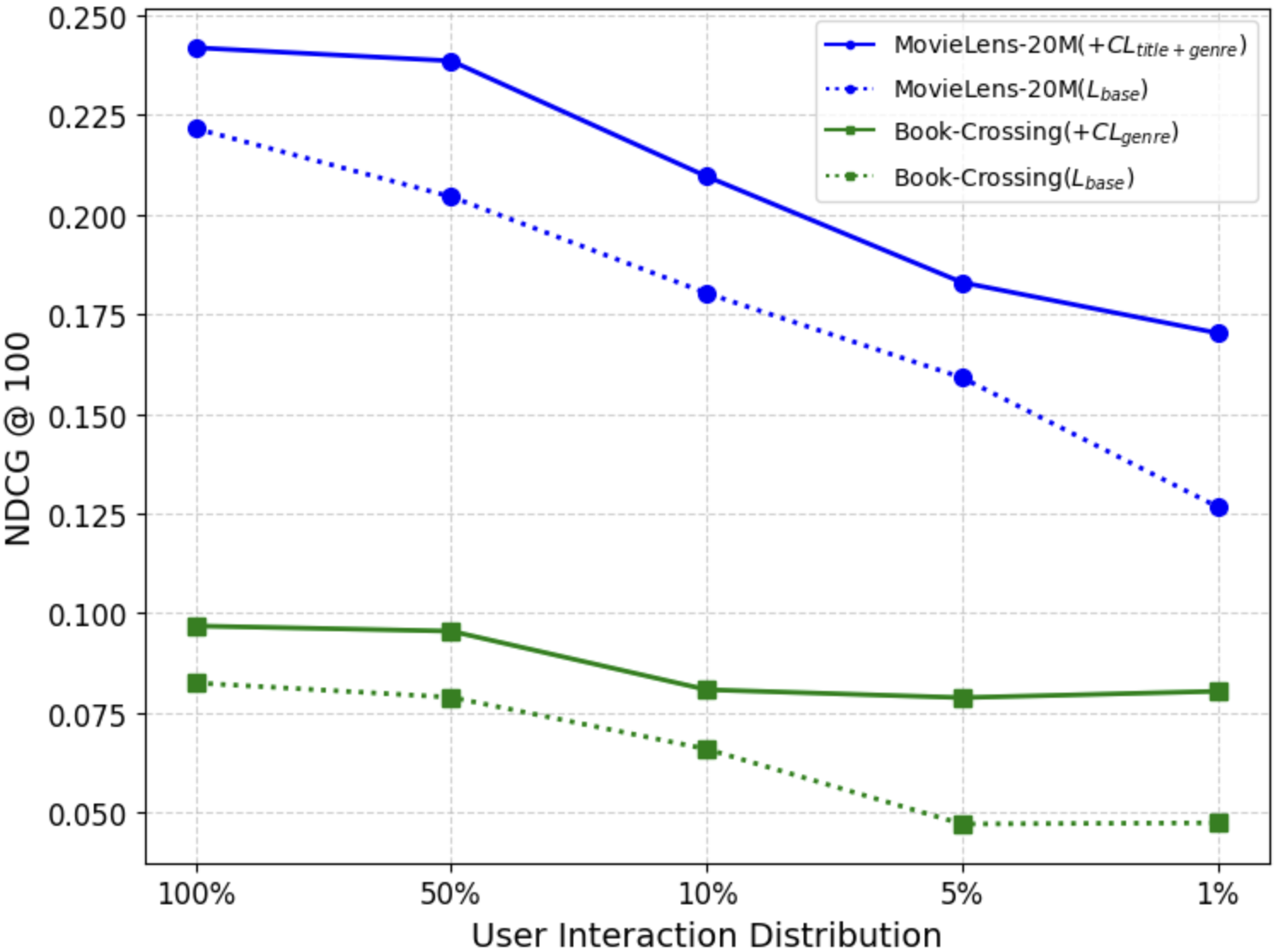}
\caption{Comparing Model Performance Across a Spectrum of User Activity Levels. The distribution on the x-axis is based on the number of user-content interactions of each test user. For example,
1\%  refers to the test users with the fewest user-content interactions.
}
\label{fig:coldstart}
\end{figure}


\begin{table*}[hbt!]
\caption{Performance of Intra- and Inter- List diversity of the proposed model compared to the baseline model. The highest performance is denoted in bold and the second best results are underlined.}
    \centering
     \resizebox{0.98\textwidth}{!}
    {
    \begin{tabular}{c|c|ccc|ccc}
    \hline\toprule
    
    \multicolumn{1}{l|}{ } & \multicolumn{1}{l|}{ } & \multicolumn{3}{c|}{\textbf{Without Semantic Text}}& \multicolumn{3}{c}{\textbf{With Semantic Text}}        \\  
        \textbf{Dataset} & \textbf{Metric} & $L_{base}$ (Baseline) & \textbf{$+ CL_{genre}$} & \textbf{$+ CL_{title+genre}$} & \textbf{\hspace{0.3cm}$L_{base}$}\hspace{0.5cm} & \textbf{$+ CL_{genre}$} & \textbf{$+ CL_{title+genre}$}  \\ \hline\toprule
        \multirow{2}{*}{Movie} &\textbf{Inter@20} & 0.7054 & 0.7478 & 0.7511 & 0.6952 & \hspace{0.5cm}\underline{0.7615} & \textbf{0.7806} \\
        &\textbf{Intra@20} & 0.1997 & 0.3427 & 0.2667 & 0.1970 & \hspace{0.5cm}\textbf{0.4407} & \underline{0.4201} \\
        \hline\toprule
        \multirow{2}{*}{Book} &\textbf{Inter@20} & 0.2963 & 0.3207 & 0.3196 & 0.3374 & \hspace{0.5cm}\underline{0.3494} & \textbf{0.3670} \\
        &\textbf{Intra@20} & 0.0048 & 0.0029 & 0.0024 & \textbf{0.0168} & \hspace{0.5cm}{0.0135} & \underline{0.0162} \\

    \hline\toprule
    \end{tabular}
    }
\label{table:diversity}
\end{table*}
\begin{figure*}
\centering 
\subfloat[Movie]{\includegraphics[width=.50\linewidth]
{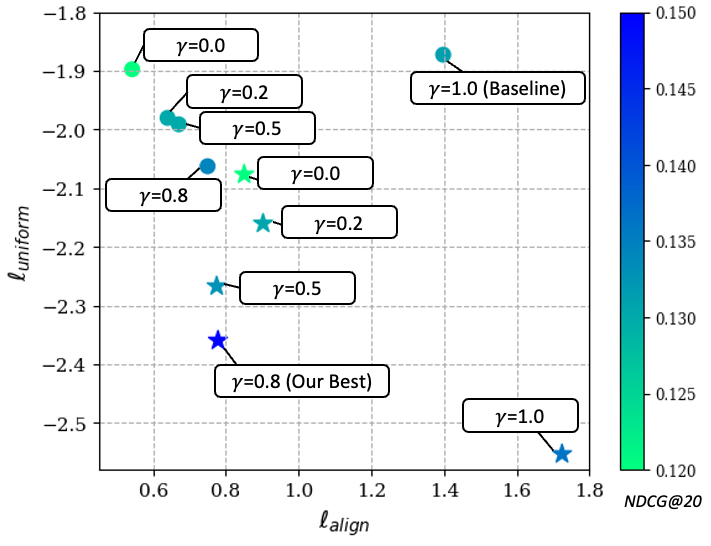}}\hspace{5pt}
\subfloat[Book]{\includegraphics[width=.47\linewidth]{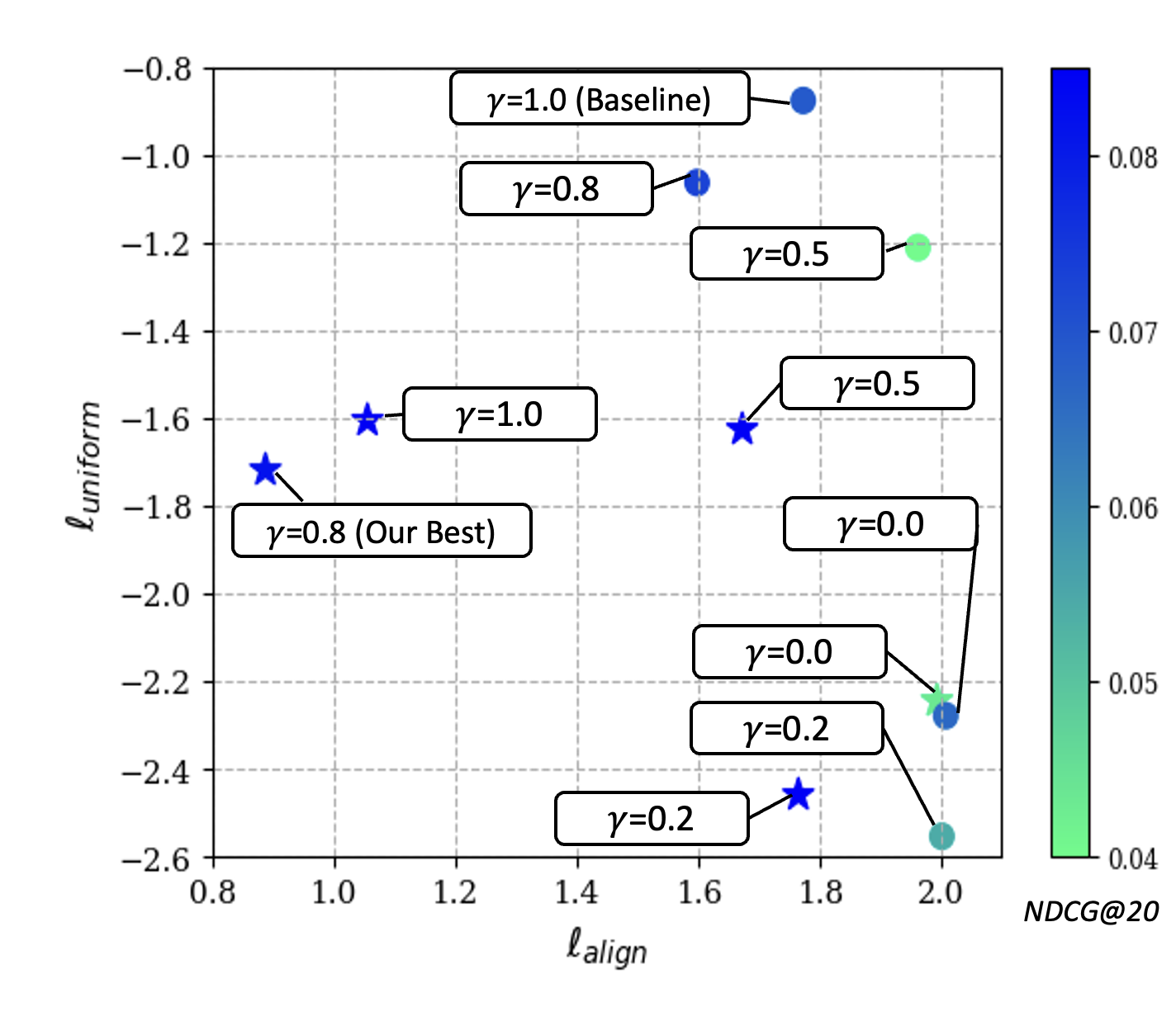}}
\caption{Performance of Embedding Quality. $\ell_{align}$-$\ell_{uniform}$ plot of models by $\gamma$ which represents a weight of base loss, $L_{base}$. Embeddings obtained from the trained model using synopsis data are marked with star points({\tiny$\bigstar$}), while circle points({\tiny\newmoon}) indicate embeddings obtained from the model trained without synopsis data. The color of points represents the NDCG@20. For both $\ell_{align}$ and $\ell_{uniform}$, lower numbers are better.}  
\label{fig:align}
\end{figure*}

\subsection{Main Results}
We present the results of our study in tables with abbreviated notation. Our baseline model is referred to as $L_{base}$\textit{(Baseline)}, trained identically to the original KGCN model \cite{wang2019knowledge1}. The `With Semantic Text' section of the tables indicates that the respective models were trained using textual input to represent the content embeddings. The notation $+CL_{(genre|title+genre)}$ denotes that a model is trained using content-based contrastive loss and $L_{base}$ jointly. $CL_{genre}$ refers to the utilization of content-content interactions extracted from genre metadata, while $CL_{title+genre}$ indicates the use of such interactions obtained from both genre and title. All results in the tables are averaged over ten random trials. 

In the CTR prediction experiment, we utilize a t-test, with each model subjected to ten trials and a comparison against the baseline model. Table \ref{table:auc} provides a summary of AUC and F1-score results. The results reveal that the models that employed our content-based contrastive loss function and semantic text information exhibited superior performance overall. The results obtained from Table \ref{table:recall_ndcg} highlight that our proposed model performs better than the baseline for all $K$ in the top-$K$ recommendation evaluation. Moreover, Figure \ref{fig:coldstart} underscores that our best model in both domains shows a lesser drop in NDCG@100 for test users with few user-content interactions compared to the baseline.

An evaluation of the diversity of recommendations in our models is performed with the findings presented in Table \ref{table:diversity}. Our proposed model, trained using $CL_{genre+title}$, achieved the highest performance in Inter-list diversity, while $CL_{genre}$ or $L_{base}$\textit{(with semantic text)} exhibited the best performance in Intra-list diversity. In the context of CTR or ranking-based performance, we found that in most cases, the use of both genre and title resulted in superior performance. However, such an approach may not always be optimal in terms of diversity. For instance, in the movie domain, if we assume that the user rates Despicable Me highly, then Despicable Me 2 might also be recommended due to a similar title. Consequently, it can boost ranking-based performance when a user consumes the recommended content. However, due to their similar embeddings, it adversely impacts the diversity performance of the recommendations. For this reason, we conclude that using $CL_{genre+title}$ scored lower diversity performance compared to $CL_{genre}$. It is imperative to clarify the Intra-list diversity result of the book domain, as they exhibit significantly lower numbers in comparison to others. This is due to the fact that each user typically engages in approximately seven user-content interactions, in contrast to the movie dataset, where there are approximately 100 interactions. Furthermore, the low Intra-list Diversity scores imply that the recommended items for each user exhibit significant similarities and this trend becomes more conspicuous when the model is trained without semantic text. This finding contributes to the examination of the subsequent embedding quality result (uniformity and alignment).
\begin{table*}[hbt!]
\caption{Comparison of performance utilizing text generated by a generative AI, LLaMA to the model using human-generated text and the baseline model. Recall@20 and NDCG@20 are respectively represented by R@20 and N@20.}
    \centering
     \resizebox{0.85\textwidth}{!}
    {
    \begin{tabular}{c|l|cc|cc}
    \hline\toprule
    
        
        \textbf{Dataset} & \textbf{Model} & \textbf{R@20} & \textbf{N@20} & \textbf{Inter} & \textbf{Intra}\hspace{0.1cm}   \\ \hline\toprule
        
        \multirow{3}{*}{Movie} & $L_{base}$ (Baseline) & 0.1967 & 0.1312 & 0.7054 & 0.1997  \\
        &$+ CL_{title+genre}$ (Human-generated text) & 0.2136 & 0.1498 & 0.7806 & 0.4201 \\
        &$+ CL_{title+genre}$ (LLaMA-generated text) & 0.2106 & 0.1504 & 0.7717 & 0.4268 \\\hline\toprule
        \multirow{3}{*}{Book} & $L_{base}$ (Baseline) & 0.0926 & 0.0690 & 0.2963 & 0.0048  \\
        &$+ CL_{title+genre}$ (Human-generated text) & 0.1041 & 0.0820 & 0.3670 & 0.0162 \\
        &$+ CL_{title+genre}$ (LLaMA-generated text) & 0.1092 & 0.0807 & 0.3437 & 0.0151 \\

    \hline\toprule
    \end{tabular}
    }
\label{table:llma}
\end{table*}
Figure \ref{fig:align} presents a comparison of the uniformity and alignment of embeddings generated by various models trained using the $CL$ loss function, based on different $\gamma$ values, along with their corresponding NDCG@20 scores. In general, models that have superior alignment and uniformity demonstrate better performance \cite{wang2020understanding}. In the movie domain, we observed that the utilization of semantic text mostly produces better uniformity in models, regardless of the value of $\gamma$. Additionally, it can be observed that as the value of gamma decreases, the alignment value also tends to decrease. If a model does not use the semantic text information, it can be inferred that the embeddings are highly anisotropic \cite{ethayarajh2019contextual} since the alignment value decreases significantly while the uniformity value does not decrease much. However, this tendency is less prominent in the book domain, aligning with our previous observations made during the examination of Intra-list diversity. This can be explained by the relatively low average number of user-content interactions and the significant similarity in item embeddings. When the user-content loss balancing parameter, $\gamma$ is set to a low value, the training process faces challenges in achieving effective convergence, as illustrated by the data points in the lower right corner of Figure \ref{fig:align} (b). Thus, by calibrating the balancing parameter $\gamma$ in accordance with the specific dataset, we can acquire well-aligned embeddings with enhanced diversity in recommendation performance. Our best model, trained with loss functions $L_{base}$ and $CL$ with a $\gamma$ value of 0.8, demonstrates robust alignment and uniformity in the experimental results.


\subsection{Ablation Studies}

In Section \ref{data}, we observed that a total of 169 synopses were missing from the movie dataset, 4,673 were absent from the book dataset, and the model training proceeded without utilizing them. It is uncertain whether this omission would have any significant impact on the overall performance of the model. In addition, it is worth noting that manual annotation of data is often a resource-intensive and time-consuming process. As such, we seek to explore whether the use of generative AI models could potentially aid in mitigating this issue. We generated 16,954 movie and 14,967 book synopses by providing a clear and concise prompt, \textit{``Briefly, a story of a (movie|book) named (movie\_title|book\_title) released (released\_year) is about "} to LLaMA\footnote{\label{llama}\url{https://ai.facebook.com/blog/large-language-model-llama-meta-ai}} \cite{touvron2023llama}. As detailed in Table \ref{table:llma}, the results indicate that the model trained using LLaMA-generated text delivers comparable performance with that of the model utilizing human-generated text, sourced from TMDB, Goodreads, and Google books with regard to diversity and ranking performance.

\section{Conclusion and Future Work}
This paper proposes a hybrid recommendation approach that uses semantic text and a Knowledge Graph (KG) in a multi-task learning framework.
We propose a novel content-based contrastive loss function that is jointly optimized with a conventional collaborative loss function. The content-based contrastive loss function utilizes a sampling strategy that creates positive and negative sets for each content based on its metadata. In this study, we aim to explore the efficacy of multi-task learning on KG neural networks that incorporate semantic text information for content recommendations. We conduct our experiments to assess the performance of the recommendation system from various perspectives. The results of our study demonstrate that the proposed model surpasses the baseline in terms of performance metrics based on ranking, diversity, and embedding quality as well as when confronted with a cold-start scenario. Through our work, we provide insights into the importance of comprehensive evaluation and analysis in obtaining valuable recommendations, emphasizing the need to consider multiple factors beyond rank-based performance metrics.

Furthermore, our methods are agnostic to the choice of model. By utilizing our proposed sampling strategy and content-based contrastive objective, any KG-based model can be employed. We performed intensive experiments on Knowledge Graph Convolutional Networks (KGCN) to demonstrate the effects of semantic embeddings and a content-based multi-task learning approach in content recommendation. Our future research will focus on two possible avenues: (1) extending the proposed methods to other KG-based neural network models, and (2) assessing the suitability of our approach within the context of real-world recommendation data.

\section*{Ethical Considerations}
Since our models were trained using the TMDB, Goodreads, and Google Books dataset or the generated text of the LLaMA, generative language model, there is a risk that they may propagate any toxic or hateful content, such as racism, insults, or xenophobia, present in the training data. To mitigate this, we suggest content moderation in preprocessing to ensure our recommendations adhere to ethical standards.

\section*{Acknowlegements}
This work was supported in part by National Science Foundation grant 212720.

\nocite{*}
\section{Bibliographical References}\label{sec:reference}

\bibliographystyle{lrec-coling2024-natbib}
\bibliography{lrec-coling2024-example}

\label{lr:ref}
\bibliographystylelanguageresource{lrec-coling2024-natbib}

\end{document}